\documentclass[reprint,twocolumn,pre,showpacs,amsmath,amssymb,aps]{revtex4-2}

\usepackage{natbib}
\usepackage{graphicx,color,rotating}
\usepackage[latin1]{inputenc}
\usepackage{textcomp}
\usepackage{dcolumn}            
\usepackage{bm}                 
\usepackage{mathtools}          
\usepackage{upgreek}
\usepackage{subdepth}  			
\usepackage{epstopdf}   		
\usepackage{float}				

\usepackage{array}
\newcolumntype{L}[1]{>{\raggedright\let\newline\\\arraybackslash\hspace{0pt}}m{#1}}
\newcolumntype{C}[1]{>{\centering\let\newline\\\arraybackslash\hspace{0pt}}m{#1}}
\newcolumntype{R}[1]{>{\raggedleft\let\newline\\\arraybackslash\hspace{0pt}}m{#1}}

\usepackage{hyperref}					
\usepackage{xcolor}						
\hypersetup{			     			
    colorlinks,
    allcolors={blue}
}
\newcommand{\etal}{\textit{et~al.}}

\newcommand{\qmarks}[1]{{``#1''}}
\newcommand{\upspace}{\rule{0ex}{2.5ex}}

\newcommand{\mr}[1]{\ensuremath{\mathrm{#1}}}
\newcommand{\myvec}[1]{\bm{#1}}
\newcommand{\ee}{\mathrm{e}}
\newcommand{\ii}{\mathrm{i}}
\newcommand{\dm}{\mathrm{d}}

\newcommand{\avr}[1]{\big\langle #1 \big\rangle}

\DeclareMathOperator{\re}{Re}

\newcommand{\iot}{{\ii\omega t}}

\newcommand{\ve}{\varepsilon}

\newcommand{\pp}{\partial}

\newcommand{\nablabf}{\boldsymbol{\nabla}}

\newcommand{\Lapl}{\nabla^2}

\newcommand{\divop}{\nablabf\cdot}

\newcommand{\scap}{\!\cdot\!}

\newcommand{\aMCF}{a_\mr{MCF\text{-}7}}

\newcommand{\CCC}{\myvec{C}}

\newcommand{\DDD}{\myvec{D}}

\newcommand{\EEE}{\myvec{E}}

\newcommand{\eee}{\myvec{e}}

\newcommand{\FFF}{\myvec{F}}

\newcommand{\FFFrad}{\myvec{F}^\mathrm{rad}}
\newcommand{\FFFpt}{\myvec{F}^\mathrm{pt}}

\newcommand{\Fptr}{F^\mathrm{pt}_r}
\newcommand{\Fptz}{F^\mathrm{pt}_z}
\newcommand{\FFFgrav}{\myvec{F}^\mathrm{grav}}
\newcommand{\Fgrav}{F^\mathrm{grav}}
\newcommand{\Fgravz}{F_z^\mathrm{grav}}
\newcommand{\FFFdrag}{\myvec{F}^\mathrm{drag}}
\newcommand{\Fdrag}{F^\mathrm{drag}}
\newcommand{\Fdragz}{F_z^\mathrm{drag}}

\newcommand{\Fradz}{F_z^{\mathrm{rad}}}

\newcommand{\fff}{\myvec{f}}

\newcommand{\kc}{k_\mathrm{c}}

\newcommand{\ks}{k_\mathrm{s}}

\newcommand{\nnn}{\myvec{n}}

\newcommand{\rrr}{\myvec{r}}

\newcommand{\rmv}{r_\mr{mv}}
\newcommand{\rmc}{r_\mr{mc}}
\newcommand{\rfl}{r_\mr{fl}}
\newcommand{\rPML}{r_\mr{PML}}

\newcommand{\uuu}{\myvec{u}}

\newcommand{\vvv}{\myvec{v}}

\newcommand{\zfl}{z_\mr{fl}}
\newcommand{\ztrap}{z_\mr{trap}}
\newcommand{\zPML}{z_\mr{PML}}

\newcommand{\calP}{\mathcal{P}}

\newcommand{\cO}{c_0}

\newcommand{\cfl}{c_\mr{fl}}

\newcommand{\kappt}{\kappa_\mathrm{pt}}

\newcommand{\kapfl}{\kappa_\mr{fl}}

\newcommand{\etaO}{\eta_0}

\newcommand{\etafl}{\eta_\mr{fl}}

\newcommand{\Gamsl}{\Gamma_\mathrm{sl}}
\newcommand{\Gamfl}{\Gamma_\mathrm{fl}}

\newcommand{\rhofl}{\rho_\mr{fl}}
\newcommand{\rhosl}{\rho_\mr{sl}}

\newcommand{\pI}{p_1}

\newcommand{\pII}{p_2}

\newcommand{\vvvI}{\vvv_1}

\newcommand{\vvvII}{\vvv_2}

\newcommand{\rhoO}{\rho_0}

\newcommand{\rhopt}{\rho_\mathrm{pt}}

\newcommand{\SIC}{\textrm{C}}

\newcommand{\SIMHz}{\textrm{MHz}}

\newcommand{\SIkg}{\textrm{kg}}
\newcommand{\SIkgm}{\textrm{kg}\:\textrm{m$^{-3}$}}

\newcommand{\SIm}{\textrm{m}}

\newcommand{\SImum}{\textrm{\textmu{}m}}

\newcommand{\SInm}{\textrm{nm}}

\newcommand{\SInN}{\textrm{nN}}
\newcommand{\SIpN}{\textrm{pN}}

\newcommand{\SIpTPa}{\textrm{TPa}^{-1}}

\newcommand{\SImPas}{\textrm{mPa}\:\textrm{s}}

\newcommand{\SIs}{\textrm{s}}

\newcommand{\SImps}{\SIm\,\SIs^{-1}}

\newcommand{\SIV}{\textrm{V}}

\newcommand{\beq}[1]{\begin{equation} \eqlab{#1}}
\newcommand{\eeq}{\end{equation}}
\newcommand{\bsub}{\begin{subequations}}
\newcommand{\esub}{\end{subequations}}
\def\bal#1\eal{\begin{align}#1\end{align}}
\def\bsubal#1\esubal{\bsub \begin{align}#1\end{align} \esub}
\newcommand{\nn}{\nonumber}
\def\bsubalat#1#2#3\esubalat{\bsuba{#1} \begin{alignat}{#2} #3 \end{alignat} \esuba}
\newcommand{\bsuba}[1]{\bsub \eqlab{#1}}
\newcommand{\esuba}{\esub}
\newcommand{\eqlab}[1]{\label{eq:#1}}
\renewcommand{\eqref}[1]{Eq.~(\ref{eq:#1})}
\newcommand{\eqnoref}[1]{(\ref{eq:#1})}

\newcommand{\figref}[1]{Fig.~\ref{fig:#1}}
\newcommand{\fignoref}[1]{\ref{fig:#1}}
\newcommand{\figsref}[2]{Figs.~\ref{fig:#1} and~\ref{fig:#2}}
\newcommand{\figlab}[1]{\label{fig:#1}}

\newcommand{\secref}[1]{Section~\ref{sec:#1}}

\newcommand{\seclab}[1]{\label{sec:#1}}
\newcommand{\tabref}[1]{Table~\ref{tab:#1}}

\newcommand{\tablab}[1]{\label{tab:#1}}

\newcommand{\sigmabf}{\bm{\sigma}}

\newcommand{\cL}{c_\mathrm{lo}}
\newcommand{\cT}{c_\mathrm{tr}}

\newcommand{\uuuI}{\myvec{u}_1}

\newcommand{\SSS}{\bm{S}}

\newcommand{\SIkPa}{\textrm{kPa}}
\newcommand{\SIVpp}{\textrm{V}_{\textrm{pp}}}
\newcommand{\etaflb}{\eta^{{\mr{b}}}_\mr{fl}}
\newcommand{\AlScNIV}{\textrm{Al}_{0.6}\textrm{Sc}_{0.4}\textrm{N}}

\begin{document}

\title{Numerical study of acoustic cell trapping above elastic membrane disks driven in higher-harmonic modes by thin-film transducers with patterned electrodes}

\author{Andr\'e G. Steckel}
\email{angust@fysik.dtu.dk}
\affiliation{Department of Physics, Technical University of Denmark,
DTU Physics Building 309, DK-2800 Kongens Lyngby, Denmark}

\author{Henrik Bruus}
\email{bruus@fysik.dtu.dk}
\affiliation{Department of Physics, Technical University of Denmark,
DTU Physics Building 309, DK-2800 Kongens Lyngby, Denmark}

\date{23 December 2021}

\begin{abstract}
Excitations of MHz acoustic modes are studied numerically in 10-$\SImum$-thick silicon disk membranes with a radius of 100 and 500~$\SImum$ actuated by an attached 1-$\SImum$-thick (AlSc)N thin-film transducer. It is shown how higher-harmonic membrane modes can be excited selectively and efficiently by appropriate patterning of the transducer electrodes. When filling the half-space above the membrane with a liquid, the higher-harmonic modes induce acoustic pressure fields in the liquid with interference patterns that result in the formation of a single, strong trapping region located 50 - $100~\SImum$ above the membrane, where a single suspended cell can be trapped in all three spatial directions. The trapping strength depends on the acoustic contrast between the cell and the liquid, and as a specific example it is shown by numerical simulation that by using a 60\% iodixanol solution, a cancer cell can be held in the trap.
\end{abstract}


\maketitle

\section{Introduction}
\seclab{intro}

Recently, the concept of thin-film-actuated devices has been introduced in the field of microscale acoustofluidics at MHz ultrasound frequencies. In 2018, Reichert \etal\ demonstrated experimentally and numerically, how to generate useful acoustofluidic responses in microchannels with a thin silicon-membrane lid driven by a lead-zirconate-titanate (PZT) thin-film transducer \cite{Reichert2018}. In this case, the transducer makes up around 15\% by volume (v/v) of the actuated membrane, which is excited while leaving the bulk part of the device inert. In 2021, we successfully modeled and experimentally validated the excitation by aluminum scandium nitrite (AlSc)N thin-film transducers of MHz-modes in millimeter-sized glass-block devices without microchannels \cite{Steckel2021}. In this system, the transducer is only 0.2\% v/v. In a follow-up study \cite{Steckel2021b}, which constitutes the theoretical foundation of our present work, we demonstrated by numerical simulation, how thin-film transducers can induce acoustofluidic responses in bulk microfluidic devices on par with those obtained by using conventional bulk PZT transducers. In Ref.~\cite{Steckel2021b}, we also studied the robustness and pointed out several advantages of using thin-film transducers: the low sensitivity of the thin-film device to the material, the thickness, and the quality factor of the thin-film transducer, and that the microfabrication techniques, by which the thin-film devices can be produced, also allow for careful shaping of the transducer electrodes, a design freedom that may be used to boost the acoustofluidic response of the device.

In this work, we follow up on the latter idea, and present a numerical study of how proper shaping of the electrodes on circular thin-film-driven membranes can increase the excitation amplitudes of the higher-harmonic vibration eigenmodes in the membrane. When placing such membranes in the bottom wall of a cavity containing a liquid, we show that specific higher harmonics of the order $n \gtrsim 10$ in the membrane induce acoustic pressure fields in the liquid with interference patterns that result in the formation of a single, strong trapping region located 50 - $100~\SImum$ above the membrane, where a single suspended cell can be trapped in all three spatial directions. The choice of this model system is motivated by the increasing use of disk-shaped membranes in acoustofluidic applications, such as capacitive micromachined ultrasonic transducers (CMUT) for imaging, inkjet printing, and testing \cite{Brenner2019}, thin-film resonators for mixing and biosensing \cite{Cui2016, Cui2019}, and silicon-membrane devices for particle manipulation~\cite{Qian2021}.

In the field of acoustofluidics, electrode shaping is of course used extensively when dealing with surface acoustic waves (SAW), as the electrodes directly define these waves \cite{Delsing2019}. However, when dealing with bulk acoustic waves (BAW), the topic of this work, electrode shaping is rarely used, and not at all for the above-mentioned membrane devices. A simple split-electrode configuration with an applied anti-symmetric driving voltage, has been used on experiments on bulk piezoelectric (PZE) transducers \cite{Bode2020, Lickert2021} and on thin-film transducers \cite{Reichert2018} to obtain a strong excitation of anti-symmetric modes for optimal particle focusing. Such systems has also been studied in numerical simulations \cite{Bora2015, Moiseyenko2019, Tahmasebipour2020, Steckel2021b}. Furthermore, in a combined experimental and numerical study, Hammarstr\"om \etal\ used a square-shaped back and a full front electrode on a bulk PZE transducer to create a dynamically-defined array of particle traps in the liquid just above the transducer \cite{Hammarstrom2021}.

In the field of microelectromechanical system (MEMS), shaping of PZE-transducer electrodes has been studied in much more detail. For example the development of energy harvesting MEMS devices involving shaping of electrodes is an active research field \cite{Wein2013, Du2017, Fu2018, Yang2018a, Luo2021}. A particular thorough study is the combined experimental and theoretical work on the excitation and detection of more than 50 permitted arbitrary modes in disc-, plate-, ring-, and beam-shaped PZT-on-silicon resonators by electrode shaping \cite{Pulskamp2012}. It is this kind of selective excitation of chosen resonance modes that we in the present work extend from solid-state MEMS to acoustofluidic devices for particle handling.

The contents of the paper is as follows. In \secref{modeling} we introduce our model system, a disk-shaped silicon-membrane-based microscale acoustofluidic system driven at MHz-frequencies by an (AlSc)N PZE thin-film transducer with patterned electrodes for selective excitation of higher-harmonic resonance modes, a system chosen for its compatibility with standard MEMS fabrication techniques. A brief summary is given of the basic theory and modeling developed in our previous work \cite{Steckel2021b}. It includes time-harmonic perturbation theory, the electromechanical theory of the elastic membrane and the linear PZE transducer, acoustics and time-averaged acoustic streaming of the liquid, boundary conditions, and the numerical implementation in the software COMSOL Multiphyscis \cite{Comsol55}. In \secref{MembraneModes}, we present the main result of the paper, the selective enhanced excitation by electrode shaping of higher-harmonic membrane resonance modes for acoustofluidic applications. In \secref{CellTrapping}, we present an application example by showing how specific higher-harmonics membrane modes may induce the above-mentioned trapping of a single suspended cell. In \secref{Discussion}, we discuss the possibility of size-dependent cell trapping and the advantages and disadvantages of the presented method of trapping, and finally in \secref{Conclusion} we present our main conclusions.

\section{Model system, theory, and numerical implementation}
\seclab{modeling}

The model system, sketched in \figref{ModelSystem}, consists of a thin disk-shaped silicon membrane completely covered by a (AlSc)N thin-film PZE transducer on one of its surfaces. The ground electrode of the transducer is always fully covering the transducer, whereas the excitation electrode may be divided into several pieces each with individual alternate-current (AC) excitation voltages. Below the membrane is air (treated as vacuum), and above it a liquid. The system is axisymmetric, so the we use cylindrical coordinates  $(r,\phi,z)$ through the paper with the $z$ axis perpendicular to the membrane through its center.

The basic theory and modeling for such a thin-film PZE-transducer-driven acoustofluidic system, was developed in a perturbation scheme involving the acoustic first-order fields and the steady time-averaged second-order fields in our previous work \cite{Steckel2021b}, founded on the theory for bulk PZE-transducer-driven systems \cite{Skov2019} taking the acoustic boundary layers into account analytically through effective boundary conditions \cite{Bach2018}. Numerical simulations based on this theory have been validated experimentally for several different microscale acoustofluidic systems \cite{Skov2019, Bode2020, Lickert2021, Steckel2021}. In the following we briefly summarize this basic theory and its numerical implementation and adapt the previous cartesian-coordinate formulation into the cylindrical coordinates of the present axisymmetric system. We do not study azimuthal variations, so the full three-dimensional (3D) problem is independent of $\phi$ and is thus reduced to a two-dimensional (2D) problem in the radial and axial coordinates $r$ and $z$, respectively.

\begin{figure}[!t]
\centering
\includegraphics[width=0.95\columnwidth]{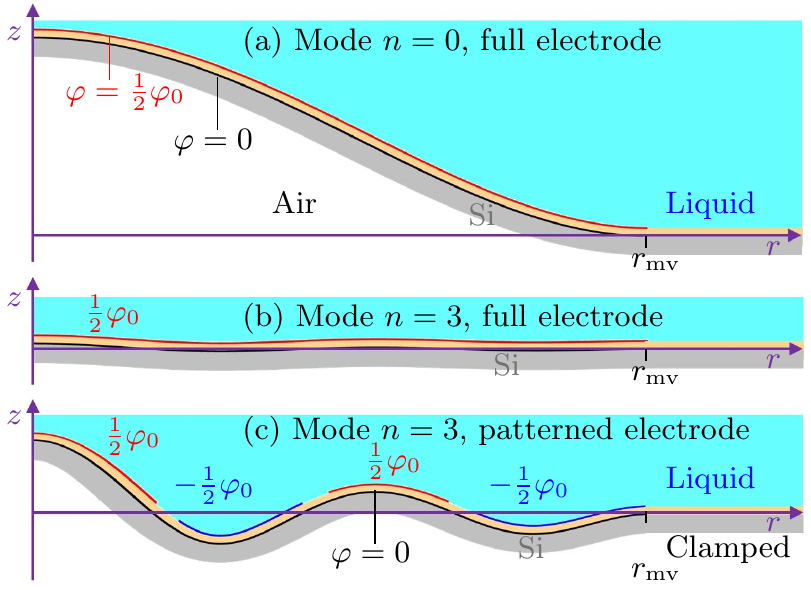}
\caption{\figlab{ModelSystem} The axisymmetric model in the $r$-$z$ plane consisting of a thin silicon membrane (gray) clamped for $r>r_\mr{mv}$ and fully covered with a thin-film (AlSc)N PZE transducer (beige) on its top surface. The grounded transducer electrode (black) covers the entire transducer, but the excitation electrode (red and blue) may be either fully covering or divided into several sections each with the indicated excitation voltage. Below the membrane is air (white, treated as vacuum) and above is a liquid (cyan). (a) The fundamental resonance mode ($n=0$) of the membrane with a large amplitude as the fully-covering excitation electrode (red) is compatible with the mode shape. (b) The third-harmonic ($n=3$) resonance mode of the membrane with a small amplitude as the fully covering excitation electrode (red) is incompatible with the mode shape. (c) The third-harmonic ($n=3$) resonance mode of the membrane with a large amplitude due to the specific patterning of the excitation electrode (red and blue) into four pieces with specific excitation voltages.
}
\end{figure}

\subsection{First-order acoustic fields}
\seclab{1st_order_fields}

Using first-order perturbation theory and complex-valued fields \cite{Steckel2021b}, the time-harmonic electric potential $\tilde{\varphi}_1(\rrr,t)$ applied to the PZE thin-film excites a time-harmonic acoustic displacement field $\tilde{\uuu}_1(\rrr,t)$,
 \beq {Time-harmonic-fields_solid}
 \tilde{\varphi}_1(\rrr,t)=\varphi_1(\rrr)\:\ee^{-\iot},
 \quad
 \tilde{\uuu}_1(\rrr,t)=\uuuI(\rrr)\:\ee^{-\iot},
 \eeq
where $\omega = 2 \pi f$ is the angular frequency, $f$ the frequency, $\rrr$ the spatial coordinate, and $t$ the time.

We convert the cartesian coordinates $\rrr = (x,y,z)$ of Refs.~\cite{Skov2019, Steckel2021b, Steckel2021} into cylindrical coordinates $\rrr = (r \cos\phi, r \sin\phi, z)$, and assuming $\phi$-independent excitation voltages, we have,
 \beq{Cyl_Fields_varphi_uI}
 \varphi_1(\rrr) =  \varphi_1(r,z),  \quad \uuuI(\rrr) = u_{r}(r,z)\eee _{r} + u_{z}(r,z)\eee _{z}.
\eeq
The governing equations of $\uuuI$ and $\varphi_1$ are the weakly damped Cauchy elastodynamic equation and the quasi-static Gauss equation for a charge-free linear dielectric,
 \beq{EquMotionSolid}
 -\rho\omega^2(1+\ii\Gamsl)\:\uuuI = \nablabf \cdot \sigmabf_{\mr{sl}}, \quad
 \nablabf \cdot \DDD = 0,
 \eeq
where $\sigmabf_{\mr{sl}}$ is the Cauchy stress tensor, $\Gamsl$ is the damping coefficient of the solid, and $\DDD = -(1+\ii \Gamma_\ve) \bm{\ve} \cdot \nablabf \varphi_1$ is the electric displacement field with $\Gamma_\ve$ and $\bm{\ve}$ being the dielectric damping coefficient and tensor, respectively. For a purely mechanical solid, the stress tensor is defined in Voigt notation as,
 \beq{StressStrainVoight}
 \sigmabf_{\mr{sl}} = \CCC \cdot \SSS_{\mr{sl}},
 \eeq
where $\CCC$ is the elastic moduli tensor of rank two which relates the strain tensor $\SSS_{\mr{sl}}$ to the stress tensor $\sigmabf_{\mr{sl}}$. In cylindrical coordinates, the Voigt notation-form of $\CCC$, $\SSS_{\mr{sl}}$ and $\sigmabf_{\mr{sl}}$ is,
 \bsubal
 \eqlab{IsotropicSolids_C_2}
 \CCC & =
 \left( \begin{array}{c@{\:}c@{\:}c@{\:}|c@{\:}c@{\:}c}
 C^{{}}_{11} & C^{{}}_{12} & C^{{}}_{13} & 0 & 0 & 0  \\
 C^{{}}_{12} & C^{{}}_{11} & C^{{}}_{13} & 0 & 0 & 0  \\
 C^{{}}_{13} & C^{{}}_{13} & C^{{}}_{33} & 0 & 0 & 0  \\ \hline
 0 & 0 & 0 & C^{{}}_{44} & 0 & 0  \\
 0 & 0 & 0 & 0 & C^{{}}_{44} & 0  \\
 0 & 0 & 0 & 0 & 0 & C^{{}}_{66}  \\
 \end{array}  \right),
 \\
 \eqlab{StrainVoigt}
 \SSS_{\mr{sl}} &=  \Big(
 \pp^{{}}_r u^{{}}_{1r}, \frac{1}{r}u^{{}}_{1r},  \pp^{{}}_z u^{{}}_{1z}, 0,
 \pp^{{}}_r u^{{}}_{1z} +  \pp^{{}}_z u^{{}}_{1r}, 0 \Big)^{\mr{T}},
 \\
 \eqlab{StressVoigt}
 \sigmabf_{\mr{sl}} &=  \Big(
 \sigma^{{}}_{rr},  \sigma^{{}}_{\phi \phi},  \sigma^{{}}_{zz},
 \sigma^{{}}_{\phi z},  \sigma^{{}}_{rz},  \sigma^{{}}_{r \phi}
 \Big)^{\mr{T}}.
\esubal
Here, the column vectors $\SSS_{\mr{sl}}$ and $\sigmabf_{\mr{sl}}$ are written as the transpose \qmarks{T} of the row vectors. Moreover, $\CCC$ is written for the lowest symmetry in our problem, the hexagonal crystal structure of (AlSc)N, and we note that it has the same form in cartesian as in cylindrical coordinates when the polarization axis is parallel to the $z$ axis. Whereas neither the displacement nor the electric field has a component in the azimuthal $\phi$-direction, the strain and stress do have a nonzero $\phi\phi$-component. Finally, we note that for silicon (111) the above form for $\CCC$ still applies, but with the symmetry constraints $C^{{}}_{13} = C^{{}}_{12}$, $C^{{}}_{33} = C^{{}}_{11}$, and $C^{{}}_{66} = C^{{}}_{44}$. Other orientations of silicon crystal are not isotropic in the $r$-$\phi$ plane and thus cannot be modeled as an axisymmetric system \cite{Hopcroft2010, Thomsen2014}.

In this work, our PZE transducer is taken to be (AlSc)N which has a hexagonal piezoelectric crystal structure \cite{Trolier2004} with the material parameters listed in \tabref{material_param}. For PZE materials, the stress and electric displacement fields are coupled to the electric and strain fields though the piezoelectric coupling tensor,
 \bsubal
 \eqlab {piezoEqu_Stress}
 \sigmabf_{\mr{sl}} &= \CCC \cdot \SSS_{\mr{sl}}  - \eee^{\mr{T}} \cdot  \EEE ,
 \\
 \eqlab {piezoEqu_Die}
 \DDD &= \eee \cdot \SSS_{\mr{sl}} +  \bm{\ve} \cdot \EEE ,
 \esubal
where the electric field defined as $\EEE = -\nablabf \varphi_1$, and where in the Voigt notation for the hexagonal crystal structure, $\eee$ and $\bm{\ve}$ are the following tensors,
\bal
 \eqlab{PiezoCoupling}
 \eee &=  \left( \begin{array}{ccc|ccc}
 0 & 0  & 0 &  0 & 0 & e^{{}}_{15} \\
 0 & 0  & 0 &  0 & e^{{}}_{15} & 0 \\
 e^{{}}_{31} & e^{{}}_{31}  & e^{{}}_{33} &  0 & 0 & 0
 \end{array}  \right)\!, \;
 \bm{\ve} =  \left( \begin{array}{ccc}
 \ve^{{}}_{11} & 0  & 0 \\
 0 & \ve^{{}}_{11}  & 0 \\
 0 & 0  & \ve^{{}}_{33}
 \end{array}  \right)\!.
 \eal

The mechanical displacement in the solids couple into the adjacent liquid or fluid as pressure waves. For a fluid with density $\rhofl$, dynamic viscosity $\etafl$, bulk viscosity $\etafl^{\mr{b}}$, compressibility $\kapfl$, and sound speed $\cfl$, and damping coefficient $\Gamfl$, the acoustic pressure $p_1$ is governed by the weakly damped Helmholtz equation, and the acoustic velocity $\vvvI^d$ is derived from $p_1$,
 \bsubalat{acoustics}{2}
 \tilde{p}_{1} & = \pI(\rrr)\: \ee^{-\iot}, \quad &
 \tilde{\vvv}_{1}^d & = \vvvI^d(\rrr)\: \ee^{-\iot},
 \\
 \eqlab {pressue_wave_p1} \Lapl \pI & =  -\dfrac{\omega^{2}}{\cfl^{2}} (1 + \ii \Gamfl)\pI , \quad &
 \kapfl &= (\rhofl \cfl^{2})^{-1},
 \\
 \eqlab {v1_fluid}  \vvvI^d & = -\ii \dfrac{1 - \ii \Gamfl}{\omega \rhofl}\nablabf \pI, \quad &
 \Gamfl &= \bigg( \frac{4}{3} \etafl + \etafl^{\mr{b}} \bigg) \omega \kapfl.
\esubalat
In our $\phi$-independent axisymmetric case we have
 \beq{Cyl_Fields_pI_vI}
 \pI(\rrr) = \pI(r,z), \quad
 \vvvI^d(\rrr) = v_{1r}^d(r,z)\:\eee _{r} + v_{1z}^d(r,z)\:\eee _{z}.
 \eeq

\subsection{Second-order time-averaged fields}
\seclab{2nd _order_fields}

The nonlinearities in the fluid dynamics induce second-order terms in the perturbation theory from products of first-order terms \cite{Steckel2021b}. Here, we focus on the time-averaged values of such fields, $\FFF_{2} = \avr{\tilde{\FFF}_{2} (\rrr , t)} = \dfrac{\omega}{2 \pi} \int _{0}^{\frac{2 \pi }{\omega}} \tilde{\FFF}_{2} (\rrr , t)\: \dm t$, in particular the steady streaming velocity $\vvvII$ and the acoustic radiation force $\FFFrad$ on suspended particles. For the time-averaged product of two first-order fields $A_1$ and $B_1$ we use the well-known expression $\avr{\re \big[ \tilde{A}_{1} (\rrr,t) \big] \re \big[\tilde{B}_{1}(\rrr,t) \big]} = \frac{1}{2} \re \big[A_{1}(\rrr) B^*_{1}(\rrr)\big]$. The acoustic streaming $\vvvII$ is an incompressible Stokes flow with a non-negligible time-average Eckart-streaming bulk force
 \beq{Second_order_gov_Equ}
 \etafl \Lapl \vvvII = \nablabf \calP_{2}
 - \dfrac{\Gamfl \omega}{2 \cfl^{2}}\re \big[ \pI^{*} \vvvI^d\big],
 \quad \divop \vvvII = 0,
 \eeq
where $\calP_{2}$ is the pressure redefined to include the excessive pressure \cite{Bach2018}. The expression for the acoustic radiation force $\FFFrad$ acting on a suspended particle of radius $a$, density $\rhopt$, and compressibility $\kappt$ is \cite{Settnes2011},
 \bsubal
 \eqlab{FRad_Eq} \FFFrad & =  - \pi a^{3} \bigg\{
 \dfrac{2 \kapfl}{3} \re\big[ f_0^{*} \pI^{*} \nablabf \pI\big]
 - \rhofl \re\big[ f_1^{*} \vvvI ^{d*} \scap \nablabf \vvvI^d\big]\bigg\},
 \\
 \eqlab{FRad_Coef_Eq} f_0 & = 1 - \dfrac{\kappt}{\kapfl}, \quad
 f_1 = \dfrac{2 (1 -\gamma) (\rhopt - \rhofl)}{2 \rhopt + (1-3\gamma)\rhofl}  ,
 \esubal
where $f_0$ and $f_1$ are the monopole and dipole scattering coefficients, $\delta$ is the width of the acoustic boundary layer, and $\gamma$ is a correction factor,
 \beq{del_gam_def}
 \delta = \sqrt{\frac{2\etafl}{\omega \rhofl}}, \quad
 \gamma = -\frac{3}{2}\bigg[1 + \ii \bigg( 1 + \frac{\delta}{a}\bigg) \bigg] \frac{\delta}{a}.
 \eeq
The total force $\FFFpt$ acting on a suspended particle moving with speed $\vvv_\mr{pt}$ at position $\rrr$ is the sum of the radiation force $\FFFrad$, the Stokes drag force $\FFFdrag$, and the buoyancy-corrected gravitational force $\FFFgrav$ from the gravitational acceleration $\myvec{g}$,
 \bsubal
 \eqlab{Fpt}
 \FFFpt &= \FFFrad + \FFFdrag + \FFFgrav,
 \\
 \eqlab{FdragFgrav}
 \FFFdrag &= 6\pi\etafl a (\vvvII - \vvv_\mr{pt}), \quad
 \FFFgrav = \frac43\pi a^3 (\rhopt - \rhofl)\:\myvec{g}.
 \esubal

\subsection{Boundary conditions}
\seclab{BC}

In the following we state the boundary conditions that we apply to the first-order acoustic and time-averaged second-order steady fields. On the axis $r=0$, we have the usual axisymmetry conditions,
 \bsub
 \begin{alignat}{4}
 \eqlab{First_order_BC_axi}
 \pp_r \pI &= 0,\;\; & \pp_r u_{1z} &= 0,\;\; & u_{1r} & = 0,\;\; & \pp_{r} \varphi_1 &= 0,
 \\
 \eqlab{Second_order_BC_axi}
 \pp_{r} \pII &= 0,\;\; & \pp_{r} v_{2z} &= 0,\;\; & v_{2r} &= 0,
 &&\text{at }\; r = 0.
 \end{alignat}
 \esub

We control the electric potential $\varphi_1$ on the transducer electrodes. At the ground electrode, $\varphi_1 = 0$ always, whereas the excited electrodes have a peak-to-peak AC voltage amplitude of $\varphi_0$, so there $\varphi = \frac{1}{2}\varphi_{0}\:\ee^{\ii \Delta}$ with $\Delta$ being a phase. Here, we work only with in-phase $\Delta = 0$ (positive) and anti-phase $\Delta = \pi$ (negative) excitation voltages, so at the electrode surfaces we have
 \bsub
 \begin{alignat}{2}
 \eqlab {BC_potentialBot} \varphi_1   &= 0, \quad & &
 \text{on ground electrodes},
 \\
 \eqlab {BC_potentialPos} \varphi_1   &= +\frac{1}{2}\varphi_{0}, \quad & &
 \text{on positive electrodes},
 \\
 \eqlab {BC_potentialNeg} \varphi_1   &= -\frac{1}{2}\varphi_{0}, , \quad & &
 \text{on negative electrodes}.
 \end{alignat}
 \esub

At interfaces away from the electrodes, we assume a zero-free-charge condition on  $\varphi_1$. On freely vibrating interfaces, we assume a zero-stress condition on $\uuu_1$,
 \beq{ZeroStresAndCharge_BC}
 \DDD \cdot  \nnn = 0,
 \quad \sigmabf_{\mr{sl}} \cdot \nnn = \bm{0},
 \quad \text{at the solid-air interface},
 \eeq

At the solid-fluid interface, we use the effective boundary conditions for the continuity of the first-order velocity and stress, developed in a coordinate-free form by Bach and Bruus~\cite{Bach2018}, where the acoustic boundary layer in the fluid has been accounted for analytically. Here we state these boundary conditions for our case with $\phi$-independent axisymmetric first-order fields,
 \bsubal
 \pp _{z} \pI   &= \dfrac{\ii \omega \rhoO}{1-\ii \Gamfl} \left [- \ii \omega u_{1z}- \dfrac{\ii}{\ks r} \pp_{r} \left (-\ii \omega r u _{1r}\right )\right ]
 \nn
 \\
 \eqlab{BC_p1} & \qquad -\dfrac{\ii}{\ks}\big(\kc^{2} \pI + \pp _{z}^{2} \pI\big),
 \\
 \eqlab {BC_u1} \sigmabf_{\mr{sl}} \cdot \eee_{z}   &= - \pI  \eee_{z}
 + \ii \ks \etaO \Big(-\ii \omega \uuuI + \dfrac{1}{\omega \rhoO} \nablabf \pI \Big),
 \\
 \kc^{2} & = (1 + \ii \Gamfl)\dfrac{\omega^{2}}{\cfl^{2}}, \qquad
 \ks = \dfrac{1+\ii}{\delta}.
 \esubal
as well as the second-order streaming $\vvv_2$,
 \bsubal
 \eqlab{v2rbc}
 v_{2r}^{d 0} &= -\dfrac{1}{2 \omega} \mr{Re} \bigg\{
 \dfrac{1}{2} v_{1r}^{\delta 0 \star}  \pp _{r}  v_{1r}^{\delta 0}
 +\dfrac{1}{2} v_{1z}^{\delta 0 \star}  \pp _{z}  v_{1r}^{\delta 0}
 \nn \\
 &\qquad\quad +    \dfrac{2-\ii}{2} \Big(\pp _{r}v_{1r}^{\delta 0 \star}
 + \frac{1}{r}v_{1r}^{\delta 0 \star} +\pp _{z}v_{1z}^{\delta 0 \star} \Big) v_{1r}^{\delta 0}
 \nn \\
 &\qquad\quad +\ii \Big(\pp _{r} u^{0 \star}_{1r} + \frac{1}{r} u^{0 \star}_{1r}
 + \pp _{z} u^{0 \star}_{1z} - \pp_{z} v_{1z}^{d \star}\Big) v_{1r}^{\delta 0}
 \nn \\
 &\qquad\quad -\ii \big(v_{1r}^{\delta 0 \star} + v_{1z}^{\delta 0 \star}\big)  \pp _{z} u_{1r}
 -  \ii \uuuI^{0 \star}\cdot \nablabf v_{1r}^{d}\bigg\},
 \\
 \eqlab{v2zbc}
 v_{2z}^{d 0} &= -\dfrac{1}{2 \omega} \mr{Re} \Big\{ \ii v_{1r}^{\delta 0 \star}
 \pp _r  u_{1z}^{0 \star} +  u_{1r}^{0 \star} \pp _{r} \big( v_{1z}^{d} + v_{1z}^{\delta 0} \big)
\nn \\
 &\qquad\quad + u_{1z}^{0 \star} \pp _{z} \big( v_{1z}^{d} + v_{1z}^{\delta 0} \big) \Big\}
\esubal
Here, $\vvvI^{\delta0} = \uuuI^0 - \vvvI^{d0}$, which together with all terms containing the factor $\ks$ originates from the boundary layer of width $\delta$.

Finally, when we consider systems without a lid, the acoustic waves transmitted from the membrane into the fluid will propagate toward infinity ($r \rightarrow \infty$ and $z \rightarrow \infty$) until damped out. To keep a finite-sized computational domain, we therefore follow Refs.~\cite{Ley2017, Skov2019} and limit the physical domain of the fluid to the region $0<r<\rfl$ and $0<z<\zfl$, then add so-called perfectly matched layers (PML) around this domain for $\rfl < r < \rPML $ or $\zfl < z < \zPML$, in which all outgoing waves from the membrane are damped out, and from which no incoming waves are sent toward the membrane,
 \bsubal
 \eqlab {BC_PML_Chifunction}
 &\text{For } \;  \rfl < r < \rPML \text{ or }\; \zfl < z < \zPML:
 \nn \\
 & \chi (r,z)=
 \\ \nn
 &K_{\mr{PML}}\bigg(\theta[r\!-\!\rfl] \bigg[\dfrac{r\!-\!\rfl}{\rPML\!-\!\rfl}\bigg]^{2}
 \!+ \theta[z\!-\!\zfl] \bigg[\dfrac{z\!-\!\zfl}{\zPML\!-\!\zfl}\bigg]^{2}\bigg),
 \\
 \eqlab {BC_PML_pp}
 &\pp_r \rightarrow [1 + \ii \chi (r,z) ] \pp _r,  \quad
 \pp_z \rightarrow [1 + \ii \chi (r,z)] \pp _z.
 \\
 \eqlab {BC_PML_dd}
 & \dm r \rightarrow \frac{\dm r}{1 + \ii \chi (r,z)},  \qquad
  \dm x \rightarrow \frac{\dm z}{1 + \ii \chi (r,z)}.
 \esubal
Here, $\theta$ is the Heaviside step function, and $K_{\mr{PML}}$ is a constant chosen, such that the outgoing waves are damped in the PML without reflections at $r=\rfl$ and $z=\zfl$.

The choice of boundary conditions at the outer surface of the PML are not crucial due to the strong damping provided in the PML. We use the Dirichlet condition $\pI = 0$ there, but changing to Neumann condition $\nnn \cdot \nablabf \pI = 0$ leads to negligible relative deviation in the result, less than $10^{-5}$ measured in terms of the L2-norm.

\subsection{Numerical implementation in COMSOL}
\seclab{Num_COMSOL}

As in Ref.~\cite{Steckel2021b}, we implement the model in the commercial finite-element software COMSOL 5.5 \cite{Comsol55}, closely following the implementation method described in
Ref.~\cite{Skov2019}. The fields $p_1$, $\uuu_1$, $\varphi_1$, $p_2$, and $\vvv_2$ are all computed with quartic-order polynomial test functions. The mesh is defined to ensure at least 20 mesh elements per wavelength, and typically more, resulting in more than 80 nodal points per wavelength in the physical domain, and 10 elements per wavelength in the PML domain. In the bulk fluid we use a triangular mesh, and in both the membrane and the thin-film transducer, a structured mesh is used with at least four mesh elements in the thickness direction. At the symmetry axis $r = 0$ is added a boundary layer of 15 elements, growing up from a smallest element of 0.005 times the bulk mesh size. This mesh ensured a successful mesh-convergence analysis similar to that presented in Ref.~\cite{Muller2014}.

\begin{table}[t]
\centering
\vspace*{-2.5mm}
\caption{\tablab{material_param} Material parameter values used in the model. The iodixanol solution is computed from the interpolation polynomials given in Ref.~\cite{Augustsson2016}. For the MCF-7 cells, the radius is chosen as the center value $10~\SImum$ in the observed range of 8.1-12.2~$\SImum$ given in Ref.~\cite{An2009}, and the scattering coefficients $f_0$ and $f_1$ are computed from  \eqref{FRad_Coef_Eq} using data from Ref.~\cite{Cushing2017}.}
\begin{ruledtabular}
\begin{tabular}{lcccc}
 Parameter &  Value &$\qquad$& Parameter    & Value \\ \hline
 \multicolumn{5}{l}{\emph{Thin-film aluminum scandium nitride},
 $\AlScNIV$ \cite{Caro2015, Olsson2020} \upspace }  \\
 $\rhosl $    & 3300 $\SIkg\:\SIm^{-3}$ &   & $\Gamsl$ & 0.0005 \\
 $C_{11}$   & 313.8 GPa  &  & $C_{33}$	& 197.1 GPa\\
 $C_{12}$   & 150.0 GPa  &  & $C_{44}$	& 108.6 GPa\\
 $C_{13}$   & 139.2 GPa  &  & $C_{66}$	& 81.9 GPa\\
 $e_{31,f}$   &$-2.65~\SIC\:\SIm^{-2}$    &
 & $e_{15}$	& $-0.32~\SIC\:\SIm^{-2}$ \\
 $e_{33}$   & 2.73 $\SIC\:\SIm^{-2}$      &
 &  $\Gamma_\ve$ & $0.0005$ \\
 $\ve_{11}$   &  22 $\ve_{0}$ &
 & $\ve_{33}$ & 22 $\ve_{0}$\\
 \multicolumn{5}{l}{\emph{Membrane silicon, Si (111)} \cite{Thomsen2014, Kim2001, Hahn2015}} \upspace \\
 $\rhosl $    & 2329 $\SIkg\:\SIm^{-3}$ &   &  &  \\
 $E$      & 168.9 GPa      &  & $s$  & 0.262           \\
 $C_{11}$ & 207.5 GPa      &  & $C_{44}$    & 73.7 GPa        \\
 $C_{12}$ & 66.9 GPa      &  & $\Gamsl$  & 0.0001  \\
 $\cL$    & 5594 $\SImps$ &  & $\cT$    & 3425 $\SImps$
 \\
 \multicolumn{5}{l}{\emph{Water} \cite{Muller2014}} \upspace \\
 $\rhofl$ & $ 997~\SIkgm$ &  & $\etafl$	 & $0.890~\SImPas$\upspace\\
 $\cfl$   & $1497~\SImps$ &  & $\etaflb$ & $2.485~\SImPas$\\
 $\kapfl$ & $448~\SIpTPa$ &  & $\Gamfl$  & $10.3~\mr{THz}^{-1} f$
 \\
 \multicolumn{5}{l}{\emph{Iodixanol 60\% solution} \cite{Augustsson2016} \upspace} \\
 $\rho_{\mr{fl}}^{\mr{Idx,}60\%}$ & $ 1320~\SIkgm$ &  & $c_{\mr{fl}}^{\mr{Idx,}60\%}$ & $ 1498~\SImps$  \upspace\\
 $\etafl^{\mr{Idx,}60\%}$ & $ 7.690~\SImPas$ &  & $\kapfl^{\mr{Idx,}60\%}$ & $ 338~\SIpTPa$
 \\
 \multicolumn{5}{l}{\emph{Cancer cell MCF-7} \cite{An2009, Cushing2017}} \upspace \\
 $\rho_{\mr{MCF\text{-}7}}$ & $ 1055~\SIkgm$ &  & $\rho_{\mr{MCF\text{-}7}}$ & $373~\SIpTPa$ \upspace\\
 $\aMCF$ & $10~\SImum$ &  & & \\
 $f_0^{\mr{Wa}}$ & $ 0.167$ &  & $f_1^{\mr{Wa}}$ & $0.037 + 0.00002\ii$ \\
 $f_0^{\mr{Idx,}60\%}$ & $ -0.104$ &  & $f_1^{\mr{Idx,}60\%}$ & $-0.153 + 0.0010\ii$ \\
\end{tabular}
\end{ruledtabular}
\end{table}

\begin{figure*}[!t]
\centering
\includegraphics[width=\textwidth]{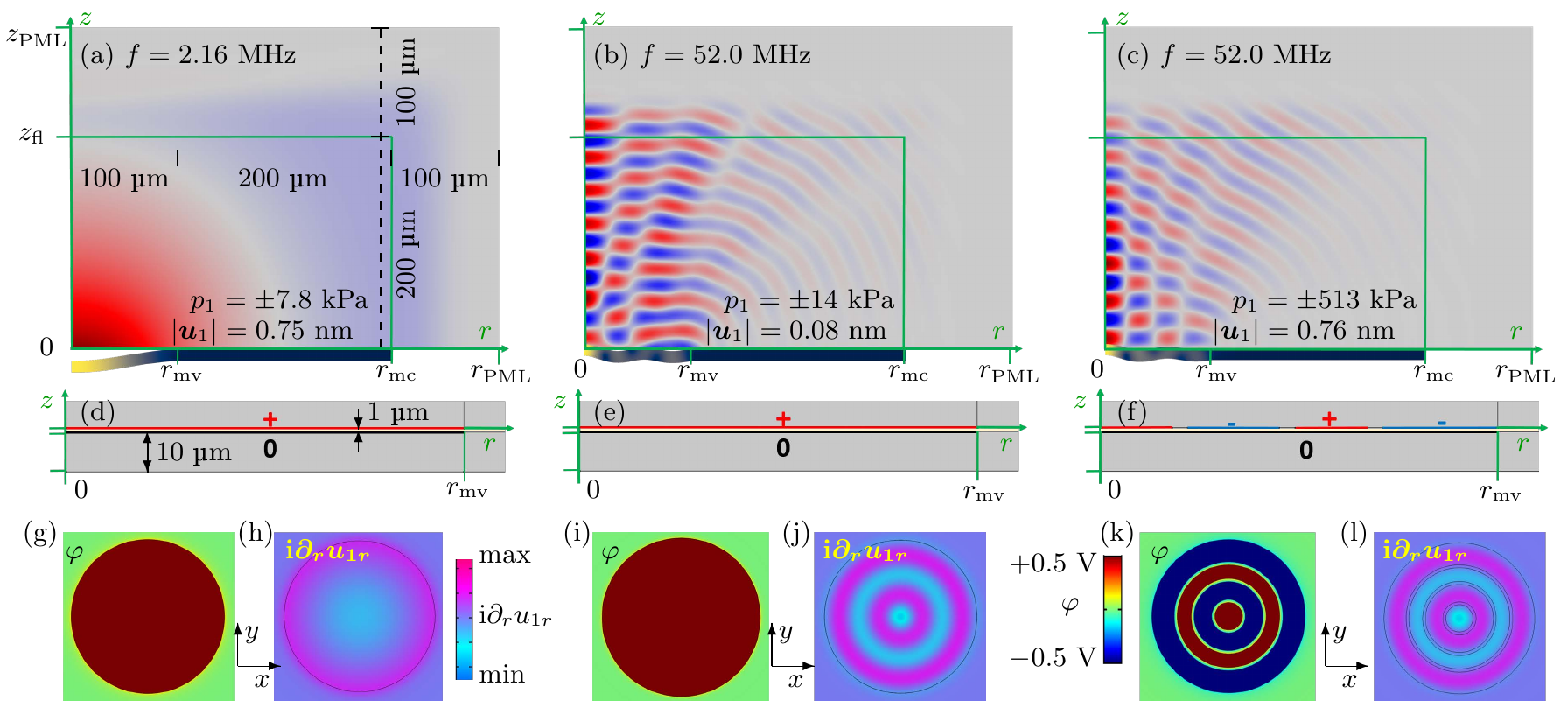}
\caption{\figlab{SmallMembrane} Simulation of the resonance modes $n$ for a 10-$\SImum$-thick silicon Si (111) membrane of radius $\rmv = 100~\SImum$ placed below a half-space filled with a fluid and actuated by an attached 1-$\SImum$-thick $\AlScNIV$ thin-film PZE transducer with a peak-to-peak AC voltage $\varphi_0 = 1~\SIVpp$. The fluid domain is surrounded by a 100-$\SImum$ wide PML domain. (a) Color plot of the pressure $\re[\ii p_1]$ from $-7.8$ (blue) to $+7.8~\SIkPa$ (red) and of the displacement amplitude $|\uuuI|$ from 0 (blue) to $0.75~\SInm$ (yellow) for the $n=0$ fundamental mode at 2.16~MHz, excited by the uniform excitation electrode shown in (d) the $r$-$z$ plane and (g) the $x$-$y$ plane. (b) The same as (a) but for the $n=3$ harmonic mode at 52.0~MHz with pressure amplitude $\pm14~\SIkPa$ and displacement amplitude 0.08~nm, excited by the uniform excitation electrode shown in (e) the $r$-$z$ plane and (i) the $x$-$y$ plane. (c) The same as (b) but with with pressure amplitude $\pm513~\SIkPa$ and displacement amplitude 0.75~nm, and with the patterned excitation electrode shown in (f) the $r$-$z$ plane and (k) the $x$-$y$ plane. (h), (j), and (l) Color plot of the out-of-phase in-plane strain $\re[\ii \pp_ru_{1r}]$ from min (cyan) to max (magenta) for the membrane modes shown respectively in (a), (b), and (c) for which the vertical displacement $\re[\ii u_{1z}]$ for clarity has been enhanced by a factor 15000, 30000, and 6000, respectively. Animations of panels (a) and (c) are given in the Supplemental Material \cite{Note1}.}
\end{figure*}

The materials used in the model are as follows: the PZE thin-film transducer is $\AlScNIV$, the silicon membrane is Si-(111), the liquid is either pure water or a 60\% aqueous solution of iodixanol, and the cell is a MCF-7 breast-cancer cell. All material parameter values used in the model are listed in \tabref{material_param} including references to the literature.

We carried out the numerical simulations of the model on a workstation with a 16-cores Intel i9-7960X processor at 3.70~GHz boost clock frequency and 128 GB of random access memory (RAM). At any given frequency, a simulation of the first-order fields $\pI$, $\uuu_{1}$, and $\varphi_{1}$ comprised of 1.4 million degrees of freedom, used 14 GB RAM, and took 50~s, whereas the second-order fields $\pII$ and $\vvvII$ comprised of 2.7  million degrees of freedom, used 27 GB RAM, and took 180~s.

\section{Excitation of higher-harmonic membrane modes by electrode patterning}
\seclab{MembraneModes}

We now introduce the excitation of higher-harmonic membrane modes in acoustofluidics. As already indicated in \figref{ModelSystem}, we show that only by patterning the excitation electrode to be compatible with the shape of the target higher-harmonic membrane mode, this mode can be excited with a sufficiently large amplitude.

\subsection{Membrane modes and patterned electrodes}
\seclab{patterned_electrodes1}

In \figref{SmallMembrane} we show the main effect of patterning the excitation electrodes appropriately for a 10-$\SImum$-thick Si-disk membrane covered uniformly by a 1-$\SImum$-thick (AlSc)N thin-film transducer. The inner part $r < \rmv = 100~\SImum$ of the membrane is free to vibrate, whereas the surrounding ring $\rmv < r < \rmc = 300~\SImum$ is clamped. Above the membrane is placed a cylindrical domain of water of radius $\rmc$ and height $\zfl = 200~\SImum$, and outside this domain is placed a PML domain of thickness $100~\SImum$.

In \figref{SmallMembrane}(a), the ground and excitation electrode are fully covering the transducer, and a peak-to-peak voltage $\varphi_0 = 1~\SIV$ is applied to the latter. A frequency sweep reveals the $\phi$-independent vibration resonance modes $n$ of the membrane disk, $n = 0, 1, 2, 3, \ldots$, with resonance frequencies $f_n = 2.16, 10.5, 19.7, 52.0$~MHz and a monotonically decreasing maximum displacement amplitude $u_{1z,n}^\mr{max} = 0.75, 0.24, 0.08, 0.02$~nm for $n = 0,1,2,3$. Animations of modes $n=0$ and $n=3$ are given in the Supplemental Material~\footnote{See Supplemental Material at \url{http://bruus-lab.dk/files/Steckel_membrane_Suppl.zip} for animations of $p_1$ and $\uuu_1$ in \figref{SmallMembrane}(a), \figref{SmallMembrane}(c), \figref{SmallMembrane3Dp1}(a), \figref{p1_lid}, and \figref{LargeMembrane}.}.

This decreasing mode amplitude is partly explained by the uniformity across the thin-film-transducer surface of the applied perpendicular voltage drop $\varphi_0$, see \figref{SmallMembrane}(d,g), which by the electromechanical coupling matrix $\eee$ promotes a uniform stretching or compression of the thin film at any given time. This works well for the fundamental mode $n=0$ at 2.16~MHz, where the film is either stretching across the entire membrane at any given time, or contracting, although not uniformly, resulting in a perpendicular displacement $u_{1z}$ that is either positive across the entire membrane, or negative, as shown in \figref{SmallMembrane}(a). For the third harmonic mode $n=3$ at 52.0~MHz, the oscillating displacement $u_{1z}$ changes sign 3 times along the radial direction as shown in \figref{SmallMembrane}(b), and this counteracts uniform stretching/contraction promoted by the uniform potential shown in \figref{SmallMembrane}(e,i), and as a result the excitation amplitude is decreased. Clearly, by arranging for a non-uniform potential following the non-uniform mechanical displacement, we could hope to restore a large excitation amplitude. But how?

The clue is the out-of-phase in-plane strain $\ii\pp_r u_{1r}$ for mode $n = 0$ and 3 shown in \figref{SmallMembrane}(h,j), respectively. This field indicates how the membrane would move were it free to vibrate in a given resonance mode. For $n=0$ at 2.16~MHz, the strain indicates a contraction (cyan) in a large center part of the membrane, which the uniform applied potential supports, only counteracting the expansion indicated by the strain in a minor peripheral part (magenta). In contrast, for mode $n=3$ at 52~MHz, the strain indicates four ring-shaped domains respectively exhibiting contraction, expansion, contraction, expansion, which is not compatible with the applied uniform potential, as each full contraction-expansion period is nearly canceled by the potential.

To make the potential for $n=3$ compatible with the strain pattern, we therefore split the excitation electrode into the same 3 ring-shaped domains (plus the central disk) defined by the strain with alternating excitation voltages $+\frac12\varphi_0$, $-\frac12\varphi_0$, $+\frac12\varphi_0$, and $-\frac12\varphi_0$, shown in \figref{SmallMembrane}(f,k). If these rings should be contacted by wires in the same layer \cite{Pulskamp2012} or by conducting vias to wires in a second layer, we leave to the experts of MEMS technology to figure out, here we just assume it possible in practise. The resulting acoustic response for the same applied voltage amplitude is shown in \figref{SmallMembrane}(c). Here we see that the resonance frequency remains $f_3 = 52.0$~MHz, and that the displacement amplitude for the patterned electrode is $u_{1z,3}^\mr{max,p} = 0.75$~nm, one and half order of magnitude higher than that of the uniform electrode, $u_{1z,3}^\mr{max,u} = 0.02$~nm, and equal to the uniform-electrode mode-0 amplitude $u_{1z,0}^\mr{max,u} = 0.76$~nm.

\subsection{Traveling and standing pressure waves induced by higher-harmonic membrane modes}
\seclab{patterned_electrodes2}

Let us now turn to the resulting acoustic pressure field $p_1$ in the liquid above the membrane vibrating in its resonance modes $n=0$ and 3. In \figref{SmallMembrane}(a) we see that a strong pressure wave is emitted from the antinode part of the vibrating membrane at 2.16~MHz in mode $n=0$ and absorbed in the PML domain. In \figref{SmallMembrane}(b), and more clearly in \figref{SmallMembrane}(c), we also see partial waves being emitted from each of the four antinodes in mode $n=3$ at 52.0~MHz. There is an important qualitative difference between the fundamental mode $n=0$ and any higher-harmonic mode $n$. The former emits only one partial wave, and thus $p_1$ is devoid of interference patterns, whereas the latter emits $n+1$ partial waves from $n+1$ ring-shaped parts of the membrane surface. These partial waves interfere and give rise to a complex interference pattern stretching out in the bulk of the liquid above the membrane, such as seen in \figref{SmallMembrane}(c). This has profound implications for the acoustofluidic properties of the system, which we investigate further in \secref{CellTrapping}.

\begin{figure}[!t]
\centering
\includegraphics[width=\columnwidth]{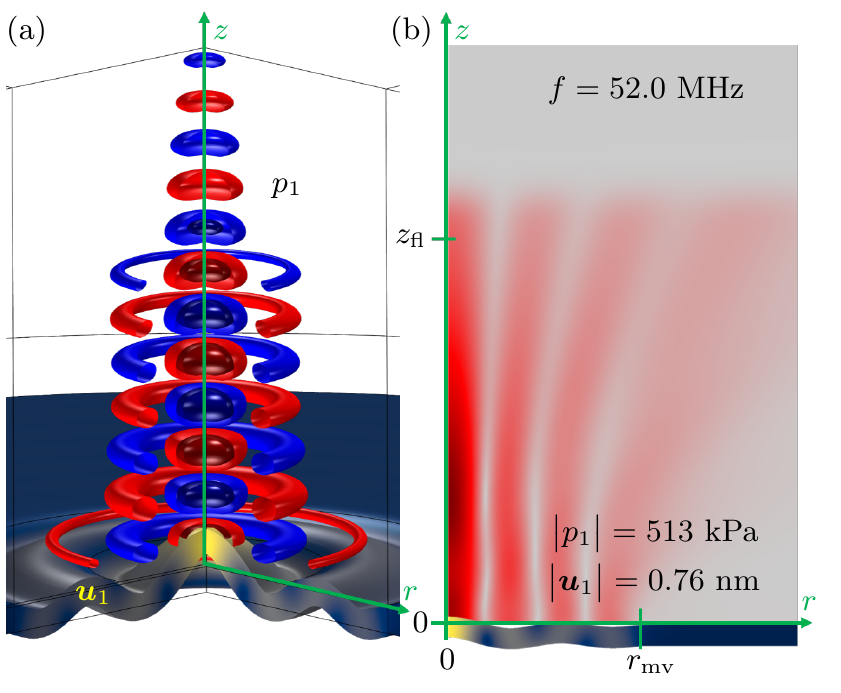}
\caption{\figlab{SmallMembrane3Dp1} (a) 3D contour plot of instantaneous isobars at $-250$, $-125$, $125$, and $250~\SIkPa$ equivalent to \figref{SmallMembrane}(c) of the acoustic pressure $p_1$ in the liquid above the membrane excited to resonance mode $n=3$ at 52.0~MHz by the patterned 3-ring electrode shown in \figref{SmallMembrane}(f,k). The displacement amplitude is enhanced by a factor 20,000 to be visible. An animation of $p_1$ and $\uuu_1$ is given in the Supplemental Material \cite{Note1}. (b) A color plot similar to \figref{SmallMembrane}(c), but for the absolute value $|p_1|$.
}
\end{figure}

In \figref{SmallMembrane3Dp1}(a) is shown a 3D contour plot of instantaneous isobars equivalent to \figref{SmallMembrane}(c) of the acoustic pressure $p_1$ in the liquid above the membrane excited to resonance mode $n=3$ at 52.0~MHz by the patterned electrode shown in \figref{SmallMembrane}(f,k). The pressure pattern consists of ring-shaped waves propagating upward in the $z$ direction, while exhibiting standing waves in the radial direction.

This particular mixture of standing radial and traveling axial waves is clearly revealed by studying the color plot \figref{SmallMembrane3Dp1}(b) of the absolute value $|p_1|$ of the acoustic pressure. Following a outward radial path from $r=0$ at any given fixed height $z = z_0$, reveals regular oscillations in $|p_1(r,z_0)|$ between nearly zero minima and large maxima, a characteristic of a standing radial wave. The most pronounced oscillation happens at height $z_0 = 60~\SImum$, and here we find the ratio $\max|p_1|/\min|p_1| \approx 140$, which corresponds to a nearly ideal standing wave.

When inspecting the pressure nodal surfaces (gray) in $|p_1|$ in \figref{SmallMembrane3Dp1}(b), we see that they form two nearly perfect and one deformed coaxial cylinder centered on the $z$-axis, but with a tendency to widen out as the distance to the membrane increases. Following any vertical path upward parallel to the $z$-axis, $|p_1|$ is mainly monotonically decreasing, which is the characteristic of a nearly ideal damped traveling wave. One exception to this behavior is found along the $z$-axis, where a local maximum in $|p_1(0,z)|$ is found for $z \approx 60~\SImum$ above the center of the membrane. Here we estimate the ratio $\max|p_1|/\min|p_1| \approx 1.3$, which corresponds to a 1.0:0.3 mixture of a traveling and a standing wave. We shall see in \secref{CellTrapping} that this small component of a standing wave along $z$ together with the large standing wave component along $r$ spawns a particle trap in all three spatial directions at some distance above the center of the vibrating silicon membrane.

\subsection{The effect of introducing a rigid lid}
\seclab{rigid_lid}

So far, we have assumed ideally absorbing surroundings by introducing the PML. As a result the pressure waves are traveling away from the membrane without being reflected back. In actual acoustofluidic systems, such a behavior is approximately realized by using soft PDMS rubber walls, as demonstrated recently by Skov \etal\ in a combined experimental and numerical study \cite{Skov2019b}. They also showed how replacing the soft lid with the other extreme, a hard glass lid, the traveling wave field was replaced by a standing wave field. Here, we briefly examine this latter case, by replacing the top PML in our model by a rigid lid placed at $z = z_\mr{fl}$ with a  hard-wall boundary condition, $\pp_z p_1 = 0$, for the acoustic pressure. We keep the PML at the side to mimic an infinitely broad system.

The results of the rigid-lid simulation of $p_1$ is shown in \figref{p1_lid}, where a vertical standing-wave behavior is clearly seen as horizontal nodal lines in the color plot of $|p_1|$ in \figref{p1_lid}(b). However, since the PML at the side still allows for the waves to propagate away from the membrane, now predominantly in the radial direction, neither the membrane modes nor the maximum pressure amplitudes are much affected by replacing the PML lid with the rigid lid, as seen when comparing Figs.~\fignoref{SmallMembrane3Dp1}(b) and \fignoref{p1_lid}(b). The smooth behavior of $|p(0,z)|$ is now overlaid with minor oscillations, but there is still a global pressure maximum along the $z$-axis.

Close to the membrane, the admixture of the vertical standing wave component is strong for the rigid-lid system, but still traveling waves are seen, as revealed by the
animations of Figs.~\fignoref{SmallMembrane}(c) and \fignoref{p1_lid}(b) given in the Supplemental Material \cite{Note1}. As the characteristics of the pressure field are not strongly affected by the lid boundary condition, we continue using the previous case of an ideally absorbing lid in the following.

\begin{figure}[!t]
\centering
\includegraphics[width=\columnwidth]{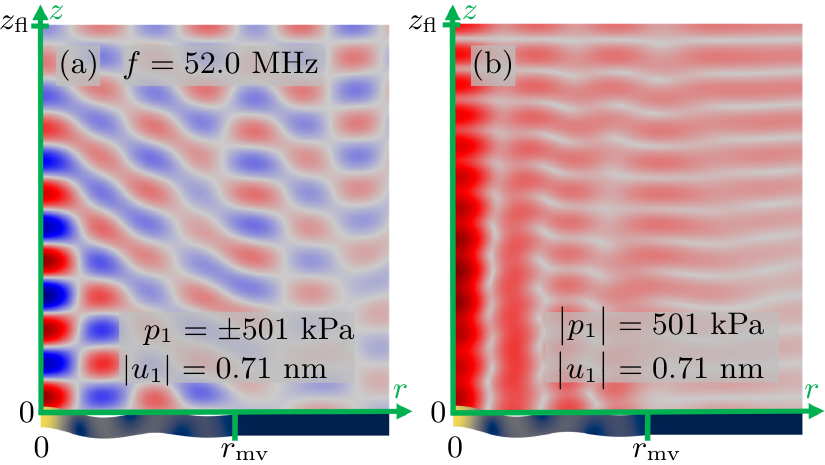}
\caption{\figlab{p1_lid} The small membrane of \figsref{SmallMembrane}{SmallMembrane3Dp1} but
with a hard-wall lid at $z = z_\mr{fl}$ instead of PML. (a) Same as \figref{SmallMembrane}(c) for the pressure $p_1$. An animation of $p_1$ and $\uuu_1$ is given in the Supplemental Material \cite{Note1}. (b) Same as \figref{SmallMembrane3Dp1}(b) for the magnitude $|p_1|$ of the pressure $p_1$.
}
\end{figure}

\section{Cell trapping by higher-harmonic membrane modes}
\seclab{CellTrapping}

We choose to illustrate the possibilities offered by the use of higher-harmonic membrane modes in acoustofluidics application with the example of trapping of a single suspended biological cell. Particle trapping is a central problem in the field studied by many groups. Examples are microparticle trapping in acoustic tweezers \cite{Baresch2016,  Karlsen2017, Riaud2017, Baudoin2019, Gong2019, Baudoin2020} and in disposable capillary tubes\cite{Hammarstrom2010, Ley2017}, nanoparticle by using seed particles \cite{Hammarstrom2014a, Evander2015, Hammarstrom2012}, and short and long term trapping\cite{Hammarstrom2014, Olofsson2021}. Trapping and the associated focusing has been created by various methods, including standing bulk \cite{Hagsater2008, Augustsson2011}, traveling bulk \cite{Bach2020}, and surface \cite{Shi2009a, Collins2016a, Skov2019b} acoustic waves. An important focus area in the field is the trapping and focusing of biological cells in general \cite{Gascoyne2002, Nilsson2009, Gustafson2021, Olofsson2021} and of circulating tumor cells in particular \cite{Li2015, Low2015, Augustsson2016}, in some cases accomplished by tuning the acoustic properties of the suspension medium tune relative to those of the given cells \cite{Augustsson2016, Olofsson2020}.

\begin{figure}[!b]
\centering
\includegraphics[width=\columnwidth]{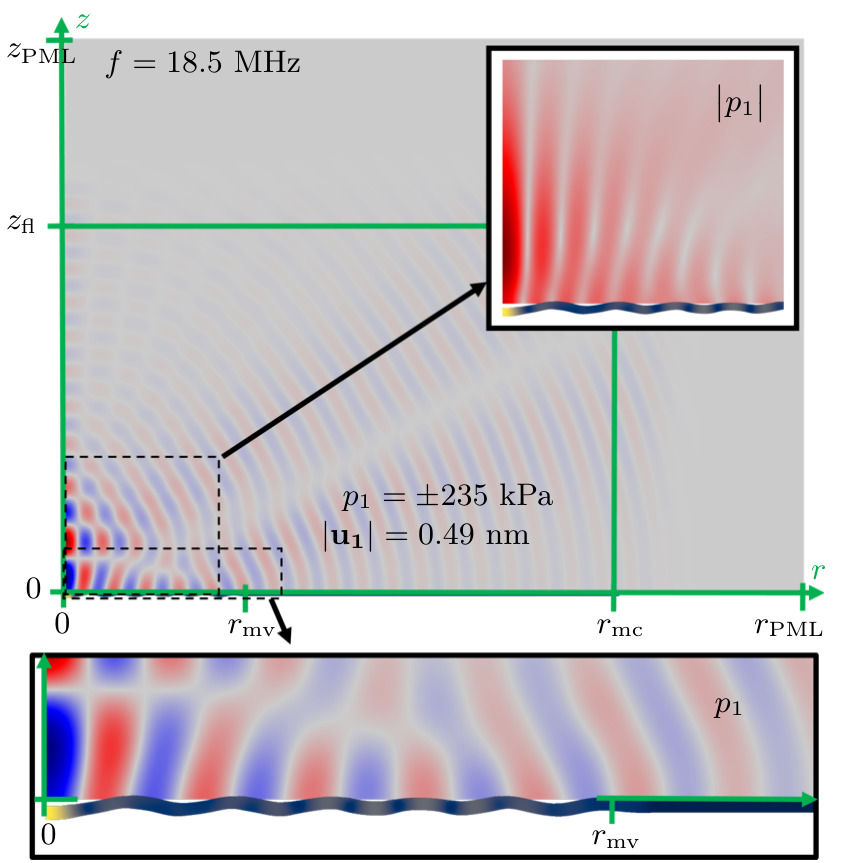}
\caption{\figlab{LargeMembrane} A membrane device as in \figref{SmallMembrane}(c), but with $\rmv = 500~\SImum$, $\rmc = 1500~\SImum$, $\zfl = 1000~\SImum$, and with the excitation electrode is divided into 10 (instead of 3) ring-shaped segments each $44~\SImum$ wide, separated by 4-$\SImum$-wide gaps (not shown), and excited with alternating voltages $\pm\frac12 \varphi_0$ as in \figref{SmallMembrane}(f,k). Color plots of the pressure $\re[\ii p_1]$ from $-235$ (blue) to $+235~\SIkPa$ (red) and of $|\uuuI|$ from 0 (blue) to $0.49~\SInm$ (yellow) for the $n=10$ higher-harmonic mode at $18.5~\SIMHz$. The top inset is a color plot of $|\pI|$. The bottom inset shows details of $\pI$ and $|\uuuI|$. An animation of $p_1$ and $\uuu_1$ is given in the Supplemental Material \cite{Note1}.
}
\end{figure}

\subsection{Design of the membrane for cell trapping}
\seclab{MembraneTrapDesign}

In the following numerical simulation analysis of cell trapping, we choose as our model cell the breast cancer cell MCF-7 with the known acoustic parameters listed in \tabref{material_param} and here assumed to be spherical with a radius $\aMCF = 10~\SImum$. To facilitate acoustic trapping of such a cell, we prefer to work with an acoustic wavelength in the fluid larger than the cell, $\lambda_0 \gtrsim 8\aMCF \approx 80~\SImum$, or $f_0 \lesssim 20~\SIMHz$, a limit in which the long-wavelength limit expression~\eqnoref{FRad_Eq} of the acoustic radiation force on a suspended particle is valid. Furthermore, for membrane mode $n$ with radial wavelength $\lambda_r^{(n)} \approx 2\rmv/(n+\frac12)$, the mode frequency scales as $f^{(n)} \propto \big[\lambda_r^{(n)}\big]^{-2}$, and since $f^{(n)} \propto \lambda_0^{-1}$ in the fluid, we can at membrane resonance write $\big[\lambda_r^{(n)}\big]^2 = \lambda_0 L$, where $L$ is a parameter of dimension length characteristic for the given system. We find numerically that $L \approx 110~\SImum$ in our system with a 10-$\SImum$-thick silicon-membrane pushing on water. Combined with the long-wavelength criterion, we obtain $\lambda^{(n)}_r =  \sqrt{\lambda_0 L} \gtrsim \sqrt{8\aMCF L} \approx 94~\SImum$. Finally, to ensure the formation of a trapping point on the $z$-axis, we must demand that $\lambda_0 < \alpha \lambda_r^{(n)}$ (where $\alpha \approx 1$ is a constant we have not been able to compute) because if not, the acoustic wave in the fluid propagates at an angle away from the $z$-axis. We thus conclude that all above criteria are satisfied if $8\aMCF \lesssim \lambda_0 \lesssim \alpha^2 L$ (equivalent to $12~\SIMHz \lesssim f_0 \lesssim 20\alpha^2~\SIMHz$) and $\lambda_r^{(n)} \gtrsim 94~\SImum$.

Consequently, as a proof of concept, we use the design shown in \figref{LargeMembrane}, consisting of a 10-$\SImum$-thick silicon Si (111) disk membrane of radius $\rmv = 500~\SImum$ driven in its $n=10$  harmonic mode with $\lambda_r^{(10)} \approx ~95~\SImum$ and an estimated resonance frequency $f^{(10)} \approx L \cO /\big[\lambda_r^{(10)}\big]^2 \approx 18.2~\SIMHz$. Similar to \figref{SmallMembrane}(f,k) of mode $n=3$, the mode $n=10$ is excited by patterning the excitation electrode of the 1-$\SImum$-thick $\AlScNIV$ transducer into ten ring-shaped segments. A $1~\SIVpp$ AC voltage is applied to the these segments with alternating phases $+\frac12 \varphi_0$ and $-\frac12 \varphi_0$, resulting in a strong excitation of resonance mode $n = 10$ at $f = 18.5~\SIMHz$, close the prediction $f^{(10)}$. All the features seen in \figref{LargeMembrane} of the disk-membrane displacement field $\uuu_1$ and of the pressure field $p_1$ in the liquid are the same as in \figref{SmallMembrane}(c), and the amplitudes are relatively large being $|\uuuI| = 0.49~\SInm$ and $|p_1| = 235~\SIkPa$. Of particular interest for trapping, we notice a local maximum in $|\pI(0,z)|$ in the inset of \figref{LargeMembrane} near $z = 100~\SImum$, similar to the one at $z = 60~\SImum$ shown in \figref{SmallMembrane3Dp1}(b).

\subsection{Characterizing the cell trap}
\seclab{CellTrap}

\begin{figure}[!t]
\centering
\includegraphics[width=\columnwidth]{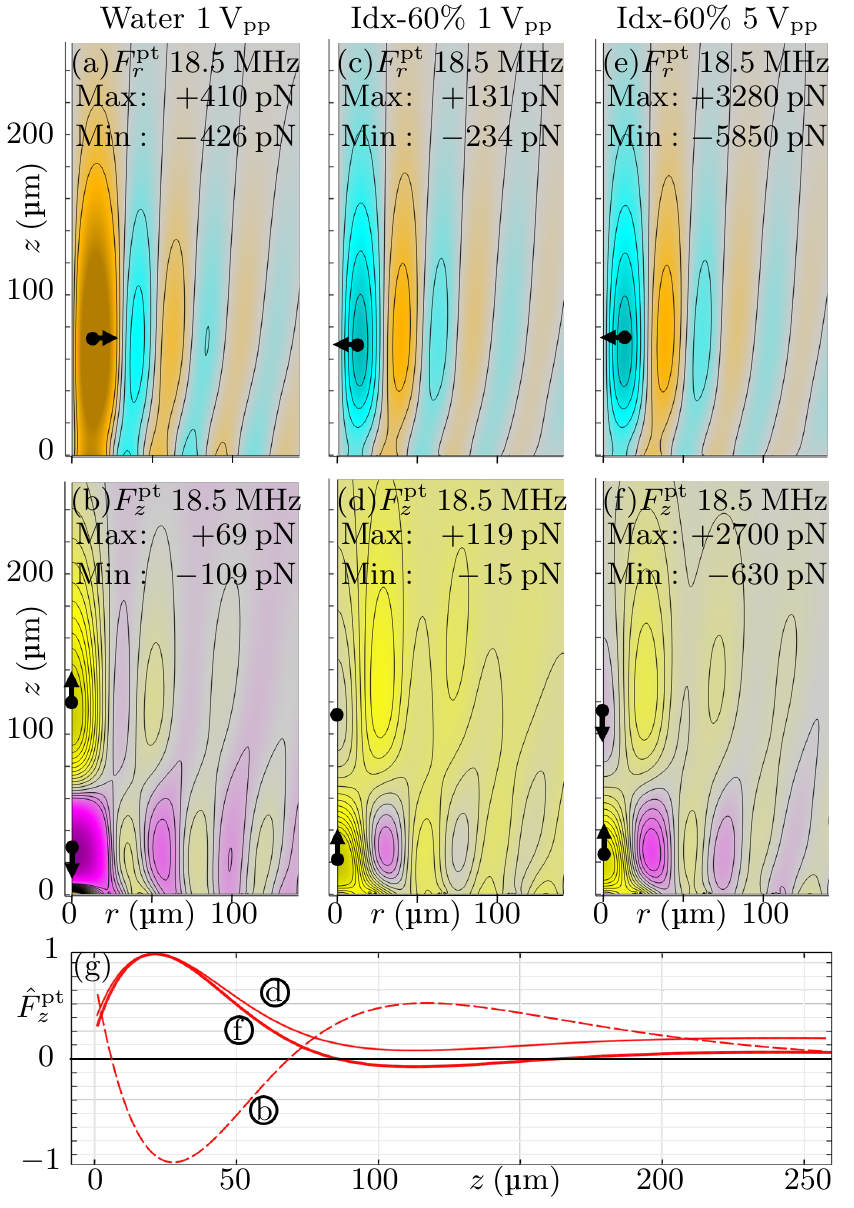}
\caption{\figlab{Frad} Contour plots of the simulated components $\Fptr$ and $\Fptz$ of the force $\FFFpt$ acting on MCF-7 cells in the system shown in \figref{LargeMembrane}, but for different liquids and voltage amplitudes: (a,b) Water at $1~\SIVpp$, (c,d) 60\% iodixanol at $1~\SIVpp$, and (e,f) 60\% iodixanol at $5~\SIVpp$, using the parameters in \tabref{material_param}. The contour lines for $\Fptr$ are from $-200$ to $100~\SIpN$ in steps of $50~\SIpN$ in (a,c), but from $-5$ to $2.5~\SInN$ in steps of $1.25~\SInN$ in (e). The contour lines for $\Fptz$ are from $-20$ to $50~\SIpN$ in steps of $5~\SIpN$ in (b,d), but $-0.5$ to $1.25~\SInN$ in steps of $0.125~\SInN$ in (f). The black arrows indicate the components direction. (g) Line plot of the normalized vertical force component $\hat{F}_z^\mr{pt}$ along the $z$-axis for case (b), (d), and (f).
}
\end{figure}


To characterize the cell trap, we study the total force $\FFFpt$ \eqnoref{Fpt} that acts on a MCF-7 cell suspended in water. In \figref{Frad}(a,b) is shown the contour plots of the radial and axial components $\Fptr$ and $\Fptz$ of $\FFFpt$. To set the scale, we note that in this case, the  buoyancy-corrected gravitational force is $\Fgrav_\mr{wa} = 2.3~\SIpN$ and the Stokes drag force is $\Fdrag_\mr{wa}(10~\SImum/\SIs) = 1.7~\SIpN$ assuming that the velocity of the cell relative to water is $10~\SImum/\SIs$. Firstly, we note that $\FFFpt$ with a magnitude above $70~\SIpN$ is completely dominated by the radiation force $\FFFrad$, secondly that $\Fptr \gtrsim \Fptz$, and thirdly that using pure water, $\FFFpt$ expels the cell from the primary nodal plane, instead of trapping it, as indicated by the black arrows.

This anti-trapping is a well-known result in trapping theory for a particle more heavy and rigid than the fluid in a single traveling wave \cite{Baudoin2020}. To change the system into a trap, we need to tune the acoustic properties of the fluid to reverse the signs of the scattering coefficients $f_0$ and $f_1$. This can be done by using the density modifier iodixanol as demonstrated in Refs.~\cite{Augustsson2016, Karlsen2016, Karlsen2018, Qiu2019}.

In \figref{Frad}(c,d) is shown that indeed by using a 60\% aqueous iodixanol solution that is heavier than pure water, the particle force $\FFFpt$ is reversed: the orange regions of $\Fptr > 0$ in \figref{Frad}(a) turns into the cyan regions of $\Fptr < 0$ in \figref{Frad}(c), and \textit{vice versa}. Also, the yellow regions of $\Fptz > 0$ in \figref{Frad}(b) turns into the magenta regions of $\Fptz < 0$ in \figref{Frad}(d), and \textit{vice versa}. The characteristic forces for the solution are $\Fgrav_{I60\%} = -11~\SIpN$ and $\Fdrag_{I60\%}(10~\SImum/\SIs) = 15~\SIpN$. In \figref{Frad}(c) we see that radial force $\Fptr$ points toward the $z$ axis for $\rmv \lesssim 20~\SImum$, reaching a maximum amplitude of 234~pN at the height $\ztrap \approx 70~\SImum$. However, the axial component $\Fptz$ is an order of magnitude smaller, and when plotting $\Fptz(0,z)$ along the $z$-axis, we find that although its has the desired upward direction $\Fptz(0,z) > 0$ for $z < \ztrap$, it is still positive at its minimum at $z = \ztrap$, $\Fptz(0,\ztrap) \approx +5~\SIpN$. The setup is thus not a trap at all, because as the numerical simulation reveals, at $\ztrap$ we have $\Fradz = -20~\SIpN$,  $\Fdragz = +14~\SIpN$, and $\Fgravz = +11~\SIpN$, such that $\Fptz = (-20 + 14)~\SIpN + 11~\SIpN = 5~\SIpN > 0~\SIpN$.

A trap is achieved simply by enhancing the AC voltage amplitude from the $1~\SIVpp$ to $5~\SIVpp$. This enhances the magnitude of the voltage-dependent second-order forces $\FFFrad$ and $\FFFdrag$ by a factor $5^2 = 25$, leaving the gravitation force unchanged. Changing from $1~\SIVpp$ to $5~\SIVpp$, we thus expect the weakest point in the trap to attain the value $\FFFpt = \big(-234\times25, (-20 + 14)\times 25 + 11\big)~\SIpN = (-5850, -119)~\SIpN$, and this is exactly what is obtained, as shown in \figref{Frad}(e,f). So using an appropriate tuning of density of the liquid and of the excitation voltage, our system is able to trap the MCF-7 cell in a trapping point located on the $z$-axis approximately $70~\SImum$ above the membrane as revealed by the red thick line in \figref{Frad}(g).

As shown in \figref{Frad}(e), radial trapping occurs once the cell enters the cyan region where $\Fptr < 0$. However, we notice that this region is surrounded by barrier-like orange region with $\Fptr > 0$, which repels the cell. This barrier could make it difficult for the cell to enter the trapping region in the first place, however once trapped, the barrier prevents more particles to enter the trap. This aspect is discussed further in \secref{Discussion}. To gain a better understanding of how the trap is loaded, we study in the steady acoustic streaming in the following.

\subsection{Steady acoustic streaming in the trap}
\seclab{Streaming}

\begin{figure}[!b]
\centering
\includegraphics[width=0.8\columnwidth]{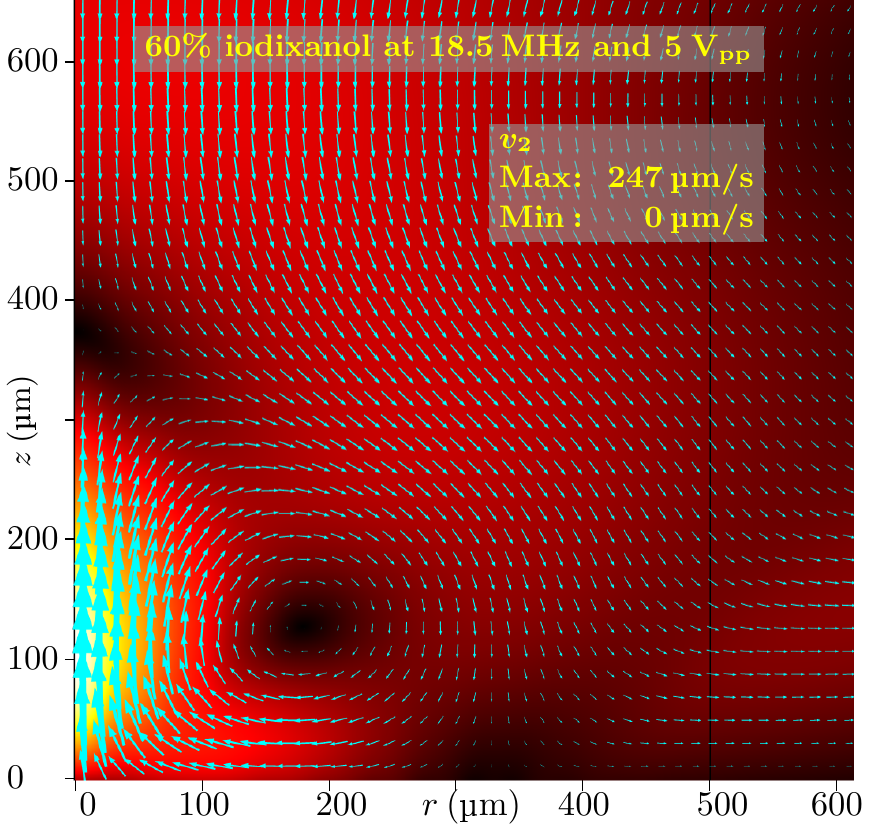}
\caption{\figlab{Streaming} Vector plot (cyan arrows) of the steady streaming $\vvvII$ and a color plot of $|\vvvII|$ from 0 (black) to $247~\SImum/\SIs$ (white) in a 60\% iodixanol solution at $f=18.5~\SIMHz$ and $5~\SIVpp$ corresponding to \figref{Frad}(e,f) of the membrane driven in mode $n=10$ . Bulk-driven Eckart streaming dominates over boundary-driven Rayleigh streaming.}
\end{figure}

In the system with an acoustic wave propagating upward from a vibrating membrane, the acoustic streaming is dominated by the bulk-driven Eckart streaming due to the force density~\eqnoref{Second_order_gov_Equ} $\fff = - \dfrac{\Gamfl \omega}{2 \cfl^{2}}\re \big[ \pI^{*} \vvvI^d\big]$ from the upward traveling acoustic wave, whereas the Rayleigh boundary-driven streaming is negligible. In \figref{Streaming} is shown the simulated streaming for the 60\% iodixanol solution at 18.5~MHz and $5~\SIVpp$ corresponding to \figref{Frad}(e,f). The dominant feature is the toroidal vortex centered at $(r,z) = (200,140)~\SImum$, which drags in particles down toward the membrane near $r = 300~\SImum$ and then sends them upward along the lower part of the $z$-axis, exactly where the trapping point is located. The flow velocity there is $247~\SImum/\SIs$ corresponding to a Stokes drag force on a trapped cell of $\Fdrag = 358~\SIpN$, which is a factor of 1.4 lower than the dominant radiation force, $\Fradz = 500~\SIpN$. The trap thus work well for the large cancer cells, but less so for smaller cells, and we can estimate the critical cell radius $a_\mr{cr}$, below which the cell is not trapped as $a_\mr{cr} = \sqrt{\Fdragz(10~\SImum)/\Fradz(10~\SImum)}\:10~\SImum = 8.5~\SImum$. \cite{Muller2012}

\section{Discussion of the results}
\seclab{Discussion}

In the following, we discuss various aspects of the higher-harmonic membrane trap, its advantages and disadvantages. Some of these aspects are particular to the specific type of trap, while other apply to single-cell traps in general.

\textit{Arrays of traps}. One distinct advantage of the membrane trap is its relatively small size, which permits the fabrication of arrays of single-cell traps for parallel analysis that can be turned on and off in a controlled manner. The design with $(\rmv,n) = (500~\SImum, 10)$ analyzed in \figref{LargeMembrane} allows for placing one trap per mm$^2$, a density that can be increased by identifying membrane modes with the desired properties for smaller values of $\rmv$.

\textit{Improved loading by turning the trap upside down}. In \secref{Streaming}, it was mentioned how the toroidal streaming vortex of the membrane trap helped loading the trap. Given that $\FFFgrav$ points upward, a further improvement of the loading process can be achieved by placing the membrane trap above the liquid instead of below. In this case cells would sediment upward toward the membrane, even when the acoustics is turned off. By subsequent actuation of the acoustics, the cells would be in a good position to be carried into the trap by the toroidal vortex. A further advantage of the inverted trap geometry is that trapping can be obtained at lower excitation voltages and for smaller particles due to a favorable interplay between $\FFFgrav$ and $\FFFdrag$ \cite{Li2021}. Turning the trap described in \secref{CellTrap} upside down reduces the critical trapping radius $a_\mr{cr}$ by 30\%.

\textit{Reduction of the radial force barrier}. If the repelling force barrier surrounding the trap region poses a problem, as mentioned in \secref{CellTrap}, it would be possible to reduce and perhaps even remove it by careful further tuning of the liquid. As studied by Qiu \etal~\cite{Qiu2019}, a range of carrier fluids with different acoustic properties can be created using solutions of Ficoll or iodixanol. In \figref{rhoTune} we show the result of a purely theoretical computation of $\FFFpt$ for the 60\% iodixanol solution of \figref{Frad}(c,d), were we assume it possible to keep all parameters constant, while lowering the density from its actual value, $\rho_{\mr{fl}}^{\mr{Idx,}60\%} = 1320~\SIkgm$, to match that of the MCF-7 cell, $\rho_{\mr{MCF\text{-}7}} = 1055~\SIkgm$, and even lower to $930~\SIkgm$. Were it possible to find a liquid with these parameters, we see in \figref{rhoTune}(a,c), how the force barrier is strongly reduced in the neutral buoyant case, and nearly vanished in the low-density case of in \figref{rhoTune}(b,d).

\begin{figure}[!t]
\centering
\includegraphics[width=1.0\columnwidth]{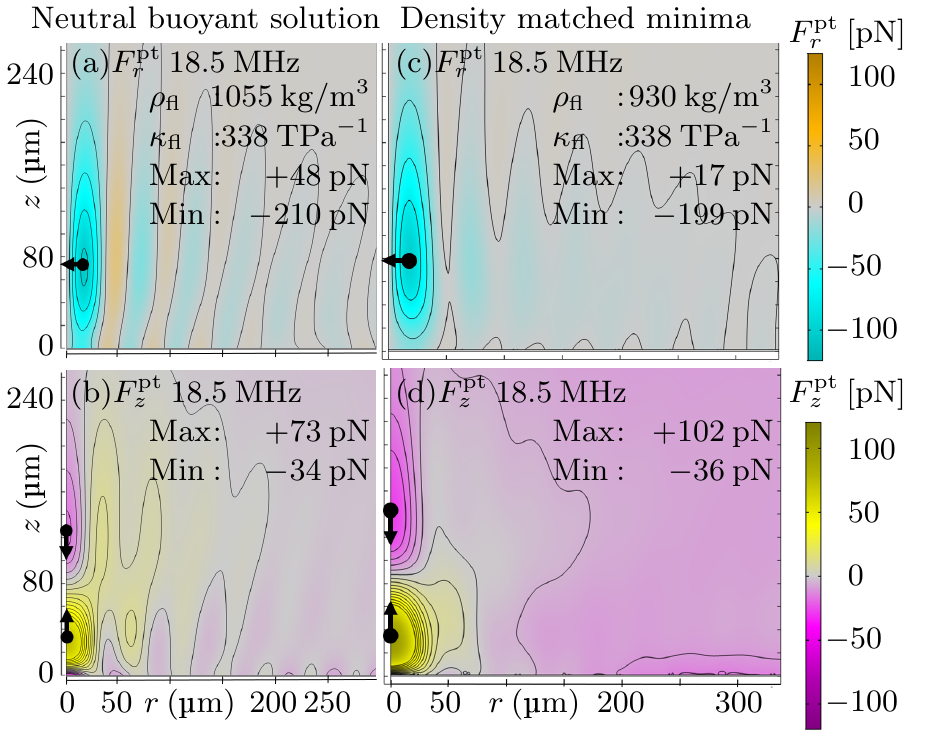}
\caption{\figlab{rhoTune} Color plots of $\Fptr$ and $\Fptz$ in a 60\% iodixanol solution at 18.5~MHz and $1~\SIVpp$ as in \figref{Frad}(c,d), but assuming theoretically a reduction in the density from its actual value $\rho_{\mr{fl}}^{\mr{Idx,}60\%} = 1320~\SIkgm$, while keeping all other parameters constant. (a) The neutral buoyant case $\rho_{\mr{fl}}^{\mr{Idx,}60\%} = \rho_{\mr{MCF\text{-}7}} = 1055~\SIkgm$. (b) A low-density case $\rho_{\mr{fl}}^{\mr{Idx,}60\%} = 930~\SIkgm$.}
\end{figure}

\textit{Creating an axial pressure node by breaking the axisymmetry}. The anti-trapping of cells in water shown in \figref{Frad}(a,b) is removed in Bessel-beam traps by breaking the axisymmetry and create a pressure node along the $z$-axis \cite{Baudoin2020}. It would be interesting to study if a similar effect could be achieved in the membrane trap. It may be done by creating segments in the excitation electrode both in the radial and azimuthal direction, and then excite a rotating higher-harmonic mode by running the excitation voltage with appropriate phase shifts in the azimuthal direction as in Ref.~\cite{Tran2012}.

\textit{Sign-change of the radiation force for shorter wave lengths}. The anti-trapping of the membrane trap may be circumvented in a different way by working in the regime where the acoustic wavelength $\lambda_0$ is comparable to the cell radius $a$. In this regime diffraction effects may change the sign of the radiation force \cite{Hasegawa1977}. A recent numerical study on suspended white blood cell has shown such a sign reversal to occur for $a \approx 0.4\lambda_0$ \cite{Habibi2017}. This points to a way of getting the membrane trap to work without tuning of the suspension medium, and would perhaps therefore also work for a standard isotonic saline solution.

\textit{Trapping of smaller-sized particles}. Given the relatively strong bulk-driven Eckart streaming in the membrane trap, it is not easy to reduce the critical radius from the above-mentioned value of $a_\mr{cr} =  8.5~\SImum$. Several of the methods proposed in the literature are for systems dominated by boundary-driven Rayleigh streaming \cite{Qiu2020, Bach2020, Winckelmann2021}. Perhaps only the method of trapped large seed particles would be a way to trap smaller particles in the membrane trap \cite{Hammarstrom2014a}.\\[5mm]

\section{Conclusion}
\seclab{Conclusion}

In this paper, we have shown in a numerical study, how the concept of selective and efficient excitation of higher-harmonic disk membrane modes can be applied to acoustofluidic systems. The physical mechanism is based on the actuation of 1-$\SImum$-thick piezoelectric $\AlScNIV$ thin-film transducers by patterning the excitation electrodes to match the out-of-phase strain pattern in the transducer at specific mode. As a proof of concept, we have shown that the $n=3$ higher-harmonic mode in an axisymmetric 200-$\SImum$-diameter, 10-$\SImum$-thick silicon membrane actuated at a frequency 52~MHz, emits pressure waves form each anti-node which results in interference patterns in the pressure field in a liquid placed above the membrane. When the condition that the wavelength in the liquid is smaller than that of the node spacing in the membrane mode, the interference pattern in the liquid can create a global pressure maximum in some distance above the membrane.

Our main example demonstrated how the global pressure maximum can be associated with the appearance of a single-cell trap, if the suspending liquid is tuned such that the cells have a negative acoustic contrast factor. In our model, this tuning was achieved by using a 60\% iodixanol solution. To lower the operating frequency and obtain a wave length larger than the cell, the $n=10$ higher-harmonic mode in 1000-$\SImum$-diameter, 10-$\SImum$-thick membrane was was excited at 18.5 MHz using appropriately patterned excitation electrodes. We showed numerically that MCF-7 cancer cells could be trapped approximately 70~$\SImum$ above the center of the membrane.  Several aspects of the trap was discussed arrays of traps, an inverted setup, reduction of the radial force barrier in the trap, the creation of an axial pressure mode, and the sign-change of the radiation force at shorter wave lengths.

Our analysis demonstrate the higher-harmonic MHz acoustic modes can be excited selectively and efficiently in acoustofluidic systems by appropriate patterning of transducer electrodes in thin-film transducers. Such modes may prove useful in practical applications, such as the single-cell trap analyzed in our main example.

\section*{Acknowledgements}

This work was supported by the \textit{BioWings} project funded by the European Union's Horizon 2020 \textit{Future and Emerging Technologies} (FET) programme, grant No.~801267.


%
%


%

\end{document}